**Electrochemistry of thin films with operando grazing incidence X-ray scattering: bypassing electrolyte scattering for high fidelity time resolved studies**

Bryan D. Paulsen,[1] Alexander Giovannitti,[2] Ruiheng Wu,[3] Joseph Strzalka,[4] Qingteng Zhang,[4] Jonathan Rivnay,[1,5*] Christopher J. Takacs[6*]

[1]Department of Biomedical Engineering, Northwestern University, Evanston, IL 60208, USA
[2]Department of Materials Science and Engineering, Stanford University, Stanford, CA 94305, USA
[3]Department of Chemistry, Northwestern University, Evanston, IL 60208, USA
[4]X-Ray Science Division, Argonne National Laboratory, Lemont, Illinois 60439, USA
[5]Simpson Querrey Institute, Northwestern University, Chicago, Illinois 60611, USA.
[6]Stanford Synchrotron Radiation Lightsource, SLAC National Accelerator Laboratory, Menlo Park, CA 94025, USA

E-mail: ctakacs@slac.stanford.edu, jrivnay@northwestern.edu

**Abstract**
Electroactive polymer thin films undergo repeated reversible structural change during operation in electrochemical applications. While synchrotron X-ray scattering is powerful for the characterization of stand-alone and ex-situ organic thin films, in situ structural characterization has been underutilized – in large part due to complications arising from supporting electrolyte scattering. This has greatly hampered the development of application relevant structure property relationships. Therefore, we have developed a new methodology for in situ and operando X-ray characterization that separates the incident and scattered X-ray beam path from the electrolyte. As a proof of concept, we demonstrate the in situ structural changes of weakly-scattering, organic mixed ionic-electronic conductor thin films in an aqueous electrolyte environment, enabling access to previously unexplored changes in the pi-pi peak and diffuse scatter in situ, while capturing the solvent swollen thin film structure which was inaccessible in previous ex situ studies. These in situ measurements improve the sensitivity to structural changes, capturing minute changes not possible ex situ, and have multimodal potential such as combined Raman measurements that also serve to validate the true in situ/operando conditions of the cell. Finally, we examine new directions enabled by this operando cell design and compare state of the art measurements.

Organic mixed ionic-electronic conductors (OMIECs) are a growing class of soft, dynamic, and functional materials with coupled mixed ionic and electronic conductivity and diverse applications across energy storage, actuators, displays, bioelectronics, and non von Neumann computing, to name a few.[1] The attractive functionality of these materials lies in their dynamic and coupled response to electrical, chemical, and mechanical stimulation. For example, in bioelectronic applications, organic electrochemical transistors (OECTs) leverage mixed conducting polymer channels to directly uptake ions to transduce and amplify biological signals.[2] Currently, design rules for OECTs and other OMIEC materials are largely absent due to a limited insights of mixed ionic and electronic transport processes. This lack of fundamental understanding is largely a failure in the structural characterization of materials under the relevant operating conditions. In particular, advances in direct structural characterization methods with molecular resolution applicable under relevant operating conditions have potential to revolutionize the development of OMIEC material design.[3–6]

High-resolution techniques like synchrotron grazing incidence wide-angle x-ray scattering (GIWAXS) have been of significant use in related applications outside of OMIEC's and have potential to give necessary insights from the molecular to mesoscale. The importance of such a test bed has analogously been shown for organic and hybrid organic inorganic perovskite photovoltaics, where the insight gleaned from in situ scattering during thin film formation has driven device advances.[7–13] Across OMIEC applications, there is a critical need to develop equivalent advanced x-ray structural characterization methods capable of high-fidelity operando measurements in relevant electrochemical environments. To date, the primary limitation has been the design of robust, x-ray compatible electrochemical cells. Specifically, the x-ray scattering and adsorption of the electrolyte, substrate, and cell components often overlaps and overwhelms the contribution from weakly ordered thin-film OMIEC materials. Aqueous electrolyte scattering gives broad features concentrated in the 1.5-3 $Å^{-1}$ range. These backgrounds can be modeled and subtracted with varying levels of success. But ultimately, advancing characterization methods and geometries that further isolate the scattering signal of interest is crucial for improving in situ/operando scattering in these systems.

Progress in the x-ray operando OMIECs studies has focused on investigations of model systems but significant compromises in fidelity and generality exist. In particular, translating these methods and insights into disordered and weakly-ordered materials is problematic. Grazing incidence x-ray scattering geometries are widely employed to increase the effective path length through thin-films and sample scattering by >100X. Additionally, x-ray total internal reflection can be used to select or suppress the scattering contribution of specific interfaces. One novel approach used an inverted structure created by transfer printing the electroactive thin film onto a thin film solid polymer electrolyte.[14] While this provides an unobstructed approach for the incident X-ray beam, the interface between electrolyte and sample failed to produce the desired total internal reflection condition. Instead, X-rays penetrate both the electroactive polymer film and the solid polymer electrolyte stack and are reflected off the underlying substrate. This reintroduces electrolyte scattering from the solid polymer electrolyte which is significant at all q's. Most importantly, the solvent-free single ion conducting polymerized ionic liquid is not representative of many aqueous application conditions as it precludes any structural effects due to electrolyte solvent or counterions, which are known to play an important role.[15–20] There has been some success in more "conventional" electrochemical cell configurations, but these required careful materials choice and significant contributions from electrolyte and/or the substrate exist.[21,22] Bischak, et. al, were able to study a highly ordered electroactive polymer film cast on the inside of metal coated Kapton window of an electrolyte filled cell. By penetrating the thin Kapton window and metal layer with the incident X-ray beam, and carefully choosing regions where the electroactive polymer scattering was accessible, operando studies were possible.[22] Similarly, in our previous work using a "cone cell" and a highly crystalline acid doped PEDOT sample, we were able to demonstrate simultaneously high-quality electrochemistry (doping and dedoping) and time-resolved scattering and diffraction.[23] This was accomplished by using a high flux, vertically focused x-ray beam (<10 µm) in a grazing incidence configuration and a sample whose thickness approached the beam height (3-6 µm). This methodology has disadvantages as multi-micron thick films are often not pertinent to desired applications. More significantly, this method does not fully overcome electrolyte background scattering for many structural features of interest (mid-to-high q features), and ultimately require relatively strong-scattering/highly-crystalline polymer films.

To address these issues, we present the "frit cell" - a new operando methodology that suppresses bulk electrolyte scattering and significantly increases the accessible q range. This methodology enables the study of disordered materials during operation with high sensitivity and fidelity. As initial proof of concept materials we selected a P-type (pg2T-TT) and N-type (p[(75:25)NDI-T2]) conjugated polymer OMIEC, both state-of-the-art mixed conductors with hydrophilic oligoethylene glycol side chains. The underlying subtle changes and the ability to study these materials under the relevant operating conditions underscore the critical need for operando studies. Validation and demonstration of multimodal capability is performed with operando Raman spectroscopy. Finally, ray tracing simulations show increased fidelity is achievable over a wide range of practical operating conditions and provide a guide for adapting and continued improvements.

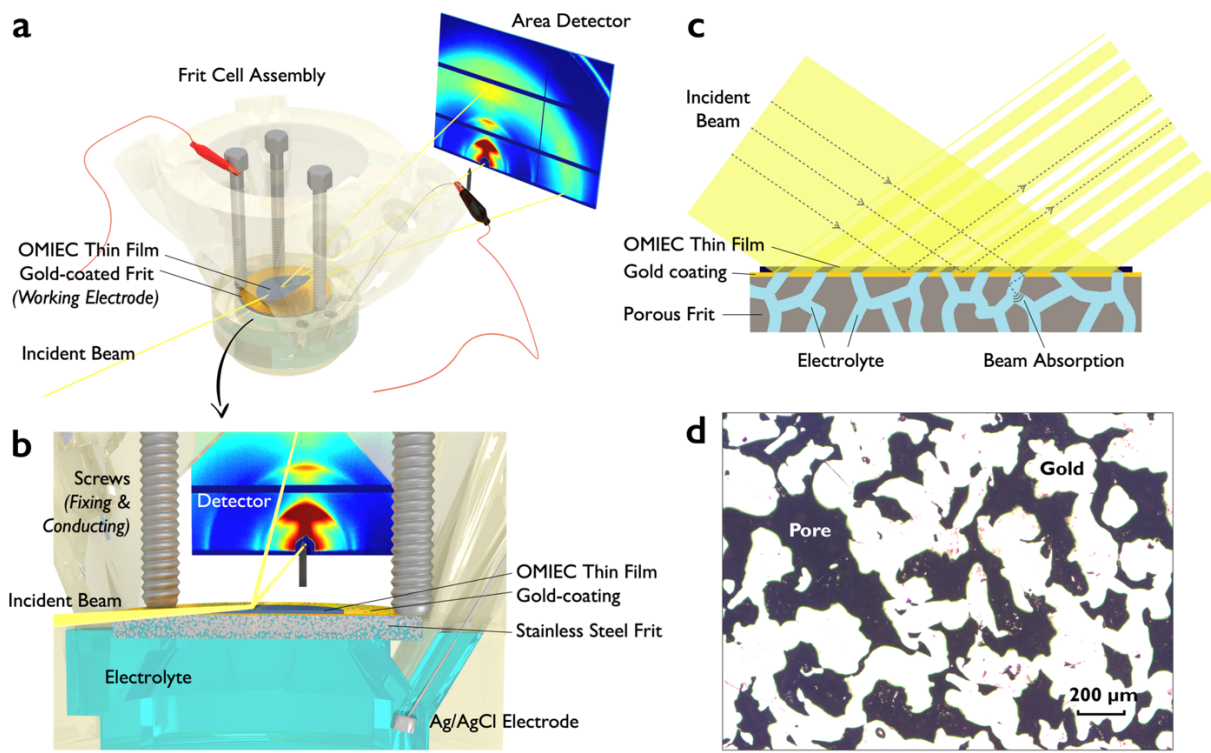

**Figure 1.** Frit Cell: (a) 3D rendering of the printed in situ cell with porous frit substrate/electrode. (b) Cross section of 3D rendering highlighting the underlying electrolyte reservoir and capillary driven electrolyte path coupling the film of interest to potentiostatic control (via an Ag/AgCl electrode in this case). (c) Schematic of the transferred film traversing the frit pore, note the frit material is a strong absorber of unreflected X-rays. The angle of incidence has been exaggerated to aid visualization. (d) Optical micrograph of the porous polished Au coated frit electrode/substrate, scale bar represents 200 μm.

Schematics of the operando frit cell, components, and its operation in grazing incidence are shown in Figure 1. The electroactive polymer thin film sample is prepared separately and transferred on top of a conductive, flat, and porous frit using established film transfer methods.[24] The result is millimeter scale continuous films that span the surface pores without rupture or collapse, as confirmed by optical microscopy before and after X-ray experiments. The electrolyte reservoir and reference/counter electrodes are located below the frit. Electrolyte is drawn into the frit through capillary forces, saturating the thin film sample. As the samples are mixed ionic-electronic conductors, lateral transport is used to equilibrate the supported and suspended regions. This lateral transport has been extensively leveraged to study ion transport.[25,26] Thin aluminum foils serve as entrance and exit x-ray cell windows and the cell head space is continually purged with water saturated $N_2$ to prevent the evaporation of water from the electroactive film. The result is unobstructed X-ray access to the thin-film sample in a grazing incidence geometry and the sample is in a hydrated state with ready access to electrolyte below.

A stainless-steel (SS) frit with a nominal 30% porosity and a manufacturer-stated 20 um pore size is used as it provides an accessible surface with nearly ideal properties for operando scattering experiments. This frit has three useful properties that combine to limit spurious background contributions to the measured signal. First, SS is a highly crystalline material and the first Bragg peak of SS is around ~2.5 Å$^{-1}$, well above the range of q's typically thought to be of interest in many polymer systems. Scattering away from the Bragg peaks is negligible resulting in a wide range of q with near zero background from the SS, an observation that we note is generally expected for a wide variety of polycrystalline metals and ceramics. Second, the top surface of the frit is polished flat using standard lapping methods and gold coated. This flat, gold coated interface allows for ohmic sample contact and creates an "x-ray mirror", i.e. total internal reflection of the incident x-ray beam for incidence angles below the critical angle of gold (~0.3 degrees). Optical micrographs (Figure 1d) show approximately 60% of the frit top surface is flat and gold coated with the rest being irregular surface pores. This flat surface portion reflects the majority of the incident beam back through the sample for a second pass, increasing the sample scattering. Third, stainless steel has a high x-ray absorbance (attenuation length of 9.7 um at 10.92 keV). Therefore, X-rays entering the electrolyte-filled surface pores have a short path length before reaching pore walls and being absorbed. The result is a strongly suppressed electrolyte background signal.

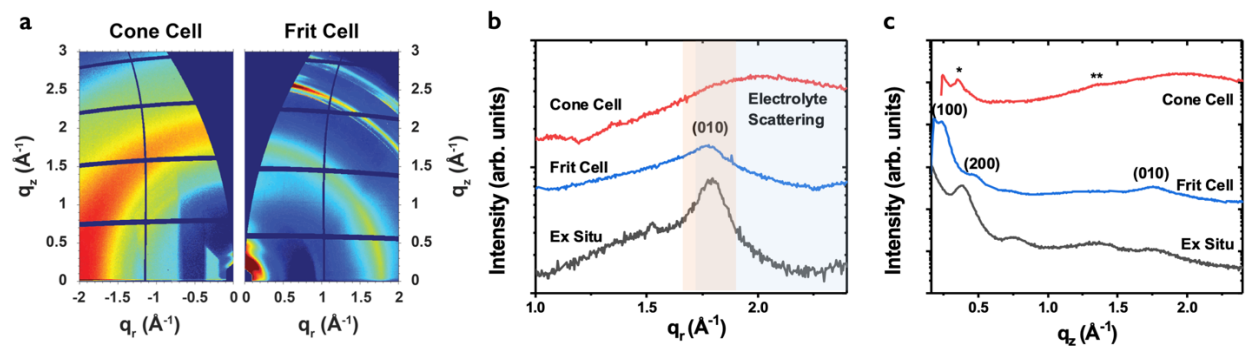

**Figure 2.** Avoiding electrolyte scattering in situ: (a) Two-dimensional grazing incidence X-ray $q_r$-$q_z$ scattering map of p(g2T-TT) (left) immersed in electrolyte in a cone cell and (right) in contact with electrolyte in the frit cell, highlighting the strong presence and absence of electrolyte scattering in the cone and frit cell, respectively. (b) In- and (c) out-of-plane line cuts from the ex situ electrolyte exposed, in situ frit cell, and in situ cone cell highlighting the difference in p(g2T-TT) and electrolyte scattering present. Peaks (*) and (**) are from the film outside the edge of the cone cell and from the PEEK cone cell itself, respectively.

Figure 2 shows the scattering of pg2T-TT, a OMIEC P-type material, collected under three conditions: electrolyte + frit cell, electrolyte + cone cell, and ex-situ samples without electrolyte. pg2T-TT is emblematic of a class of well-performing, weakly scattering materials with only a few, broad features. Using our previous cone cell design, successful in the study of more crystalline materials, the significance (and difficulty) of electrolyte scattering is well illustrated. Features like the (010) π-stacking peak are practically unresolvable (Figure 2b), lost within the electrolyte scattering. With the frit cell, the diffuse electrolyte scattering feature is absent and both the (010) π-stack scattering and multiple orders of (h00) lamellar scattering of pg2T-TT are readily discernible. Compared with the ex situ samples previously exposed to electrolyte, the main scattering features are all qualitatively present. This highlights the fidelity of the frit cell and the low backgrounds that will be achievable in operando measurements.

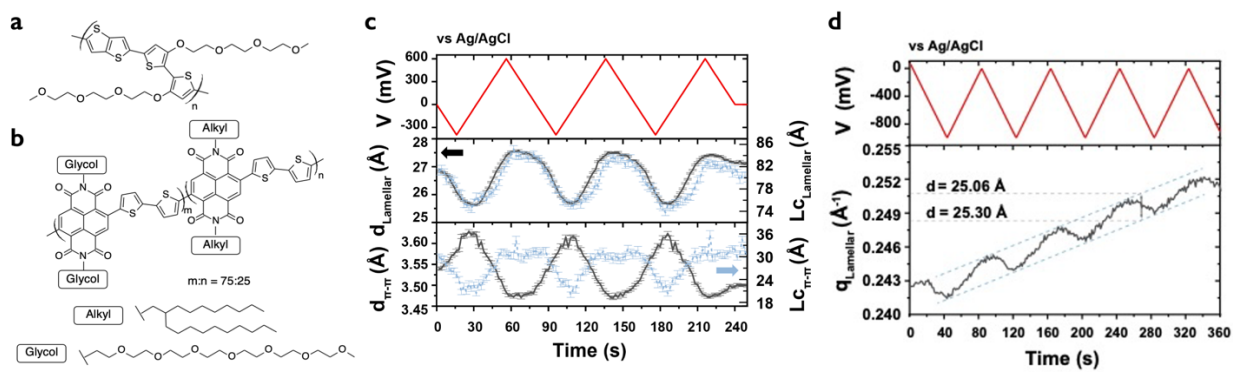

**Figure 3.** Chemical structures of (a) p(g2T-TT) and (b) p[(75:25)NDI-T2]. (c) The p(g2T-TT) lamellar/pi peak center and coherence length ($L_C$) over time during swept electrochemical potential under a hydrated nitrogen atmosphere. (d) The p[(75:25)NDI-T2] lamellar peak center over time during swept electrochemical potential in air, highlighting the small d-spacing changes resolvable with the frit cell that are not ascertained with ex situ measurements.

Most obviously, the frit based in situ measurement reveals that ex situ grossly underestimates lamellar expansion. The (h00) lamellar peaks are strongly shifted to lower q when measured in situ. Assuming a simple model where the electrolyte swell crystalline lamella, this would suggest that ex situ measurements fail to capture an approximately 10 Å lamellar expansion that coincides with electrolyte exposure.[27] Such an expansion is most likely the result of significant electrolyte uptake into the pg2T-TT crystallites although more complex polymer rearrangement in the presence of electrolyte could be possible. Regardless, these observations further underscore the need for in situ and operando studies. This effect is not simply due to water, as the hydrated pg2T-TT films do not show so large an expansion. Instead, this effect is ascribed to the compound effect of water and ions when the film is in contact with an electrolyte.

At least for aqueous systems, this erodes the geometric/volumetric argument from d-spacings that has been invoked to support the hypothesis that dopant ions preferentially reside in the amorphous regions.[14,28] Given the suggested volume expansion of the lamella, water and ions likely reside within the crystallites in both the doped and dedoped state. Further, this highlights the limits of solvent-free in situ and ex situ systems such as ionic liquids (ILs) and polymerized ionic liquids (PILs).[14,28] While they have paved the way for these results, being solvent free (molten), and in the case of PILs, single ion conductors, they are not informative prototypical systems to base our structural understanding of electroactive thin films in more traditional organic or aqueous electrolytes.

pg2T-TT appears to show some structural hysteresis during electrochemical sweeps. Figure 3a shows the time-dependent estimated d-spacings and crystallite coherence lengths - an approximate measure of crystallite size. We note these features correspond primarily to the crystalline features within the sample. While it is too early to be conclusive, the results here indicate that similar to molecular doping,[29] ion uptake required for electrochemical doping may not disrupt polymer ordering, but enhance it (compared to the electrolyte exposed de-doped state). The lamellar d-spacing and coherence length change by 10% during the operando experiment. Interestingly, these values depend not only on potential but also sweep direction, i.e. d-spacings and coherence lengths at a fixed potential are different when transitioning from dedoped-to-doped (dV/dt is positive) and doped-to-dedoped (dV/dt is negative). As a function of time, the ratio of the lamellar coherence length to peak position is nearly constant except for an apparent lag sweeping from de-doped-to-doped. This implies crystallites are almost able to uniformly contract but that expansion may proceed locally through a stepwise mechanism (e.g. Type II diffusion).[30,31] This hysteresis is more apparent in the π-stacking peak near the 0 V crossings. The π-stack d-spacing shifts by nearly 0.1 Å with the close packed structure having a 60% larger coherence length. A simplistic interpretation suggests this expanding the crystal by a few additional chains. The apparent "memory" of the system is generally interesting as well and we speculate there may be some useful connections with the modeling of nonequilibrium chemical systems along with more traditional areas of polymer theory for transport in glassy polymers.[30–32] While all of the above is deserving of follow up studies, these results highlight the need for high precision operando measurements, and the nuanced information available with the frit cell.

Importantly, operando measurements enable significant improvements in resolution and confidence. In measurements of ex situ samples, variations due to sample alignment limit the confidence in small changes and in turn limit the practical sensitivity. Figure 3b illustrates the sensitivity and fidelity achievable with in situ measurements of p[(75:25)NDI-T2], a well-performing N-type material. Often, a peak shift of 0.005 Å$^{-1}$ would be considered insignificant, and reporting such small changes in the literature would likely be met with skepticism or disregarded. Moreover, changes in diffuse scattering beyond π-stacking would be of questionable relevance (Figure S2). This operando cell appears to largely address these issues. Sub second time resolution is achievable (Figure S3) if necessary and many of the practical limits appear to be beamline stability. For example, previous ex situ investigations of a variety of napthalenediimide bithiophene backbone polymers have arrived at the conclusion that in many cases the lamellar and π-stack d-spacings do not change with electrochemical cycling.[33–35] However, the in situ electrochemical potential cycling of p[(75:25)NDI-T2] revealed a consistent potential dependent lamellar fluctuation of less than 0.25 Å (superimposed on a long time scale continuous expansion). We again note this peak is likely monitoring changes in ordered fractions of the film and not necessarily accounting for changes in more disordered/amorphous states. However, as both this technique and beamline stability improve, it may be possible to track broader, more subtle changes that correlate with disordered material. Thus, this frit based operando test bed dramatically improves the sensitivity of GIWAXS to capture minute (but not necessarily unimportant) structural changes.

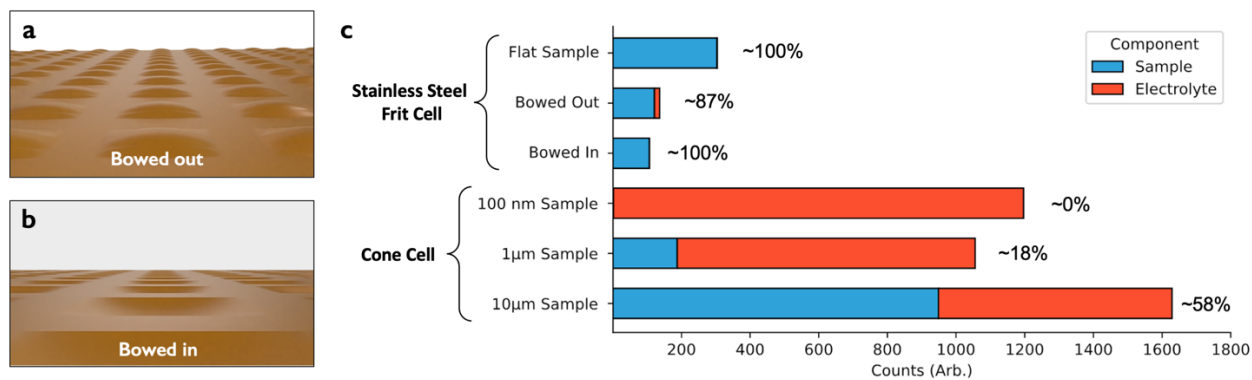

**Figure 4.** 3D geometries for thin-films a) bowing outwards and b) bowing inwards over model frit pores. The frit pores are 50 μm X 50 μm on a 70 μm grid. c) Results of an X-ray ray tracing calculation where the measured scattering is separated into the components. Percentage values are the scattering sensitivity as defined as the component of the entire scattered intensity recorded at the detector due to sample scattering. The simulated beam is a gaussian profile (3 μm x 200 μm beam) with a 0.15 degree grazing incidence. The sample thickness is 1 μm unless otherwise specified. The electrolyte contribution is significantly suppressed in the frit cell geometry compared to our previous work with the cone cell.

Here we use simulations to examine how sample swelling may change the scattering geometry due to exposure to vapor, electrolyte, and/or electrochemical cycling and how the frit cell compares to our previous cone cell design. While swelling is often present on flat electrodes and can in extreme cases lead to cracking and delamination, the frit geometry with a suspended membrane responds differently, accommodating expansion through bowing, Figure 4a and b. We developed a Monte Carlo X-ray ray tracing tool that uses 3D models of the cell, sample, and its components. This allows us to compare the sample elastic scattering to the electrolyte and other cell components for different geometries and materials parameters in silico. Figure 4c shows sample bowing indeed changes the contributions of both electrolyte and sample (Simulations use a 3 µm vertically focused beam and a 1 µm thick sample). Compared to a flat film, bowing (5 µm maximum displacement modeled here) decreases the overall scattered intensity by moving a portion of the sample outside the incident X-ray path, either by bowing in and being shadowed by the frit substrate or bowing out with a portion of the sample above the simulated incident beam height. In the case of the flat and bowed in film, electrolyte scattering is negligible. When bowed out, additional electrolyte scattering is introduced. Still, the situation is quite manageable. This is all consistent with our observed high fidelity; however, it does point out that the overall scattering intensity can change if the bowing occurs during an experiment. The overall background level is substantially higher in the cone cell where ~80% of the total scattering arises from the supporting electrolyte. The 39% decrease of scattering intensity from sample film in the in the cone cell (compared to the flat frit cell condition) arises from electrolyte absorption of the incident and scattered photons. This leads to a non-monotonic dependency of sample scattered intensity on film thickness. Increasing sample thickness in the cone cell by a factor of ten (to 10 µm) yields less than a five-fold increase in sample scattered intensity reaching the detector. More crucially, decreasing sample thickness in the cone cell by a factor of ten (to 100 nm) results in exceptionally low signal levels in qualitative agreement with our experimental results. The vertical beam size, sample thickness, and signal-to-background are all strongly linked for the cone cell. By comparison, the frit cell will tend to be more forgiving and have low background levels, allowing it to be extended to thinner samples.

As a general method, there is still considerable room for improvement, particularly in regards to the quality of the electrochemistry. While the polished, gold-coated stainless-steel frits have considerable advantages for scattering, the electrochemical side-reactions are significantly higher compared to the cone cell. Initial work in galvanizing/passivating the internal pores is promising but exploring a wider range of materials will likely be fruitful. In particular, metallic-coated, porous ceramics are likely to be promising, particularly for corrosive electrolytes.

As validation of the cell electrochemistry (electrochemical doping and de-doping) and demonstration of multimodal characterization, operando Raman experiments were demonstrated in the frit cell and compared to conventional designs (Supporting Information). Beyond the multimodal incorporation of Raman, the frit cell presented here should be compatible with other reflectance mode spectroscopies, and microscopies (optical, fluorescent, scanning probe, etc). Together, the adaptability of the Frit cell to other synchrotron techniques

(such as grazing incidence small angle scattering) and the wide range of multimodal characterization combinations opens exciting new opportunities for in situ and operando characterization.

While developed for in situ and operando scattering for OMIECs, the frit cell has great potential in a wide variety of applications in catalysis due to the easy availability of a liquid-solid-vapor interface and membrane sciences where structure is sensitive to the hydrated state. Further, this methodology is potentially applicable for measuring ion concentration in OMIECs through attenuation of the reflected beam, fluorescence from ionic species, and even potentially operando depth-resolution through grazing exit fluorescence measurements.

In conclusion, we report a new cell design for electrochemistry with in situ and operando grazing X-ray scattering measurements that allows for the time resolved measurements of weakly scattering materials by isolating the electrolyte from the X-ray path. This new cell design is based around a porous frit that acts both as the substrate for the thin film of interest and the current collector. Employing this frit cell, we capture the potential dependent d-spacings and coherence lengths of a polymeric OMIEC. Operando measurements reveal the dramatic effects of electrolyte swelling not captured by ex situ measurements. Additionally, by measuring the same location over time, in situ measurements resolve much smaller potential dependent structural changes than equivalent ex situ methods. The multimodal compatibility of the frit cell was leveraged to carry out in situ Raman spectroscopy. Finally, Monte Carlo simulations give a quantitative assessment of the benefit gained by separating the supporting electrolyte from the X-ray path. Beyond providing a general test bed for application relevant operando structural characterization of OMIECs, this frit cell design opens whole new areas of in situ and operando studies of thin film electroactive materials, across fields and applications.

**Acknowledgments**
B.P., R.W., and J.R. gratefully acknowledge support from the National Science Foundation Grant No. NSF DMR-1751308. A.G. acknowledges funding from the TomKat Center for Sustainable Energy at Stanford University. This research used resources of the Advanced Photon Source, a U.S. Department of Energy (DOE) Office of Science User Facility operated for the DOE Office of Science by Argonne National Laboratory under Contract No. DE-AC02-06CH11357. Use of the Stanford Synchrotron Radiation Lightsource, SLAC National Accelerator Laboratory, is supported by the U.S. Department of Energy, Office of Science, Office of Basic Energy Sciences under Contract No. DE-AC02-76SF00515. This work made use of the Keck-II, NUFAB, and SPID facilities of Northwestern University's NUANCE Center, which has received support from the SHyNE Resource (NSF ECCS-2025633), the IIN, and Northwestern's MRSEC program (NSF DMR-1720139). This work made use of the MatCI Facility supported by the MRSEC program of the National

Science Foundation (DMR-1720139) at the Materials Research Center of Northwestern University. Thanks to Ross Arthur, Tim Dunn, Bart Johnson, Hans-Georg Steinrueck, Kevin Stone, Vivek Thampy, Chris Tassone, and Mike Toney for useful discussion; and Quentin Thiburce and Xudong Ji for assistance with the gold deposition.

## Supporting Information

**Methods**

*Cell fabrication.* Cells were designed using Autodesk Fusion 360 and sliced in Ultimaker Cura. The cell used at APS was printed on an Ultimaker S5 with Ultimaker polycarbonate filament and a 0.25mm nozzle. This cell was designed to fit on the same mount as the cone cell and accommodate 0.8" diameter stainless steel frits (IDEX). The SSRL operando Cell was printed from 3DX-TECH PEKK-C on an Intamsys Funmat HT. The printing parameters were optimized to achieve watertight construction in both cases. The APS cell was sealed with entrance and exit windows were 8 um aluminum foils (MTI Corporation) as entrance and exit windows, affixed with Kapton tape. The scattering from these windows falls primarily within the Bragg peaks and no detectable contribution is observed within the wavevectors of interest. Kapton tape was used to seal openings on the top and side that also provide optical access during alignment. The SSRL cell was unsealed and operated under ambient conditions.

*Frit substrate/electrode preparation.* Stainless steel frits with 20 μm average pore size were mounted with wax on a sample block and successively wet polished with P600, P800, and P1200 grit sanding paper on a polishing wheel. This was followed by additional polishing with 1 μm diamond suspension and 0.06 μm alumina suspension. A 5 nm Ti adhesion layer and 100 nm Au layer was deposited on the polished surface of the frit with e-beam deposition. Samples prepared for work at SSRL were lapped by hand.

*Film transfer.* pg2T-TT was dissolved in chloroform (5mg/ml) and solutions were drop cast on Si wafer previously coated with spin on Teflon (for stamp transfer) or a sacrificial polyelectrolyte (for float transfer). Drop cast films were allowed to slowly dry in a solvent saturated environment. For stamp transfer, the films were then transferred with a PDMS stamp from the Teflon coated Si to the Au coated polished frit. Transfer to the Au coated frit was carried out on a hot plate set to 40° C. For float transfer preparation of p[(75:25)NDI-T2], Glass substrates were cleaned by sonication for 15 mins in DI water, acetone and IPA. A solution of Polydama:TFSI[36] (10 mg/mL) in acetonitrile was prepared and stirred at 55 °C for 30 mins (sacrificial layer material). The solution was spin cast onto the glass substrate (1000 RPM, 1 min), followed by the deposition of the redox-active material (p[(75:25)NDI-T2]) by spin coating (1000 RPM, 1 mins). Optical microscopy confirmed the integrity of the transferred film suspended over the frit pores. Example optical micrographs of the as cast p[(75:25)NDI-T2], the frit substrate, and the resulting float transferred film are shown Figure S1.

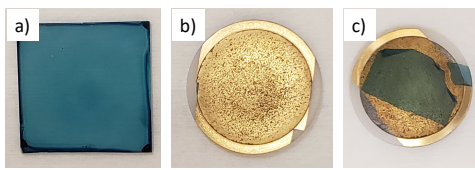

**Figure S1.** Optical micrographs of p[(75:25)NDI-T2]) samples. a) Substrate containing the interlayer and p[(75:25)NDI-T2]) b) Gold coated frit c) Transferred polymer film.

*Grazing incidence wide angle X-ray scattering.* GIWAXS measurements of pg2T-TT were performed at beamline 8-ID-E of the Advanced Photon Source, Argonne National Laboratory with 10.92 keV synchrotron radiation, 200 μm wide, ≈10 μm high, impinging on the sample with an incident angle of 0.14° (selected to be between the critical angle of the polymer and Si or Au/steel substrate), producing a 4 mm long beam foot print. Data for the frit cell and cone cells were collected in different experimental campaigns. The beam was vertically focused at the sample by a Be compound refractive lens composed of 17 lenslets with radius 0.2 mm positioned 2.175 m upstream of the sample. For the cone cell, the beam was vertically focused to a height <10 μm at the sample by a Be compound refractive lens composed of 17 lenslets with radius 0.2 mm positioned 2.175 m upstream of the sample,[23] resulting in a 4-fold increase in flux and an 8-fold increase in flux density, relative to the unfocused beam used in the frit cell measurements, where slits limited the beam to 10 μm vertically. Experiments with the cone cell were conducted within a larger helium chamber to minimize the air scatter background. Experiments with the frit cell were conducted in air with lead tape used to block upstream air scatter from reaching the detector. All measurements were at room temperature with images collected by a Pilatus 1M pixel array detector 228.2 mm away from the sample.

The cone cell was formed by pressing a machined knife edge polyether ether ketone (PEEK) cone against the pg2T-TT sample on a Si substrate supported by a polished quartz disk in a custom lens tube (Thorlabs) assembly and then filled with 100 mM aqueous NaCl electrolyte and sealed with a PTFE coated silicone septum. The frit cell was formed by placing the polished Au coated frit substrate/working electrode with the transferred film on interest into the 3-D printed in situ cell. The frit was held in place by three stainless steel screws, with which electrical contact was made. 100 mM aqueous NaCl electrolyte is added to the reservoir below the frit. The upstream and downstream beam path passed through 8μm thick Al foil windows. The internal volume of the frit cell was sealed with Kapton tape, and continually purged with water saturated $N_2$. In both cells a Ag/AgCl pellet (Warner Instruments) reference/counter electrode was employed. Potential control was carried out with a potentiostat (Ivium).

Frit cell experiments for p[(75:25)NDI-T2] were performed at SSRL BL10-2 at 12.0 keV using a focused beam with a nominal height of 50um and a Eiger 1M detector. Potential control was carried out using a Bio-Logic potentiostat. All experiments at SSRL were performed in air and the angle of incidence was varied between 0.1 to 0.3deg.

*Ray Tracing Simulations.* The Monte Carlo ray tracing was performed using a custom written Python code. 3D models of substrate, electrolyte, and sample used in the simulations were created in Fusion360 and exported as separate STL files. The 'trimesh' and Embree python libraries were used to import the 3D volumes and propagate/compute ray/object intersection. The X-ray optical constants for each material were computed using the 'xrayutilities' library. A monte carlo sampling technique was used to generate primary and scattered rays accounting; the scattering rate was set by the total elastic scattering cross-section. Without loss of generality, the scattering angles were fixed (20 degrees) for the calculations herein. A simple model for complete total internal reflection below a fixed critical angle was added at the sample/substrate interface. The sensitivity is defined as the ratio of the scattering events between the sample and electrolyte. Jupyter notebooks and model STL files are available in through a public github repository (https://github.com/ctakacs/Raytracing-FritCell-Paper) for those interested in exploring additional sample configurations.

*Raman Spectroscopy.* The operando Raman measurements were taken with HORIBA LabRam confocal microscope system. The exciting source were a HeNe laser (Melles Griot, 2.5mW) at 633 nm focused by a ×10 objective lens. The data acquisition time for each frame was 1s. For the measurement of pg2T-TT on frit cell, the upper support section of the cell (marked as translucent in the figure 1a) above the frit plane was removed to allow the focusing of objective lens. For the measurement of pg2T-TT on ITO glass, a PDMS well was used for holding aqueous electrolyte. The 633 nm laser was focused through the electrolyte (approximately 3 mm) and onto the top surface of the polymer thin-film. Potential control and current measurement were carried out by using a two-electrode setup with a potentiostat (Ivium). Ag/AgCl electrode was used as the counter/reference electrode, and gold or ITO was used as the working electrode. All data analysis and plotting were executed with MATLAB manuscript.

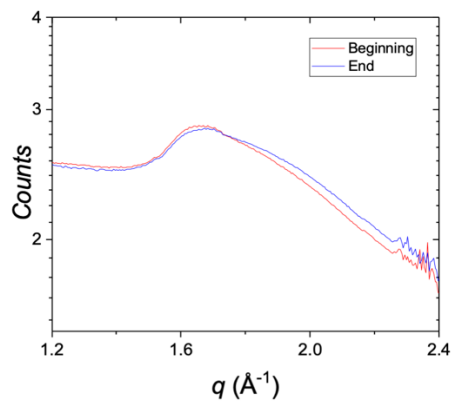

**Figure S2.** pgNDI-T2 in situ GIWAXS: out-of-plane line cuts of p[(75:25)NDI-T2] highlighting the subtle changes in π-π stack and diffuse scattering before and after potential cycling, highlighting the sensitivity to subtle structural change available with the frit cell.

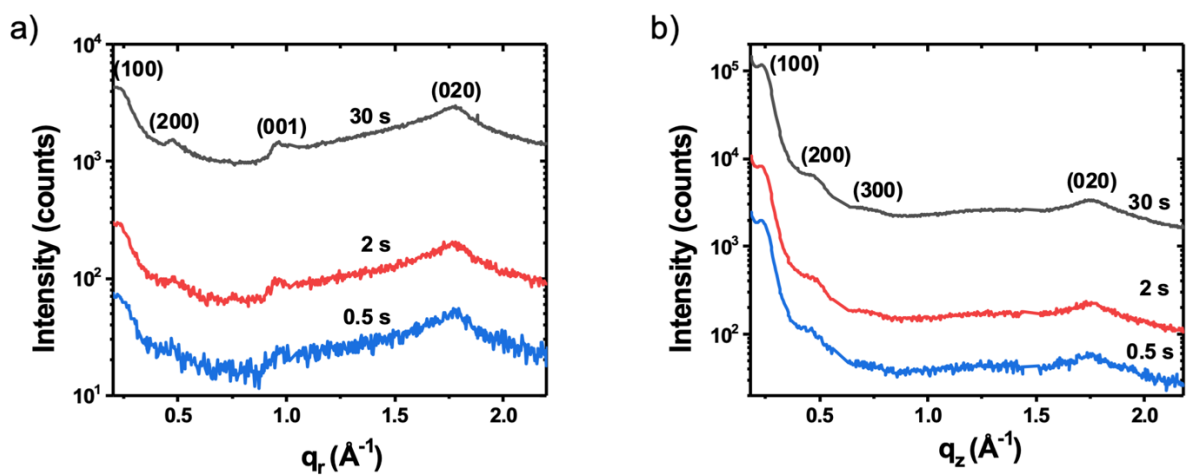

**Figure S3.** pg2T-TT in situ GIWAXS: (a) in-plane and (b) out-of-plane line cuts with varied X-ray exposure time.

**Frit Cell Based Raman Spectroscopy**

To validate and complement the x-ray results, we collected operando Raman spectra of pg2T-TT using the frit based in situ cell and compared with the same measurements carried out on a more traditional ITO coated glass working electrode, Figure S3. At 633 nm, the neutral species was resonantly excited, which provides accurate characterization of the concentration and structural changes of neutral species in the doped/dedoped pg2T-TT. The Raman peak of dedoped pg2T-TT was identified by previous DFT calculations of pg2T-TT analogs.[37] The peak at 1405 cm$^{-1}$ corresponds to the C=C stretching on thienothiophene ring (mode A), whereas the peak at 1449 cm$^{-1}$ (mode B) and 1494 cm$^{-1}$ (mode C) corresponds to the intra-ring C=C stretching and conjugated C=C/C-C stretching localized on thiophene ring, respectively. After doping, the C=C stretching peak on thienothiophene ring did not shift significantly, while both stretching peaks on thiophene showed pronounced red shift (Figure S3, mode B to 1445 cm$^{-1}$ and mode C to 1492 cm$^{-1}$), consistent with the previous Raman study of P3HT.[38] As the concentration of neutral species decreased after doping, the intensity of Raman signal was considerably weaker than for the dedoped state. The results on both ITO and frit cell showed the same trend in the peak intensity and position during operando experiments, which confirmed the successful film doping while using the frit cell. The Raman signal obtained from the frit cell had a larger peak width and a weaker S/N ratio due to the unevenness of the film caused by the voids on the frit cell surface. This both confirmed the adaptability of the frit-based cell for multimodal characterization, and independently validated the pg2T-TT charging (which is difficult to ascertain amperometrically due to steel frit side reactions).

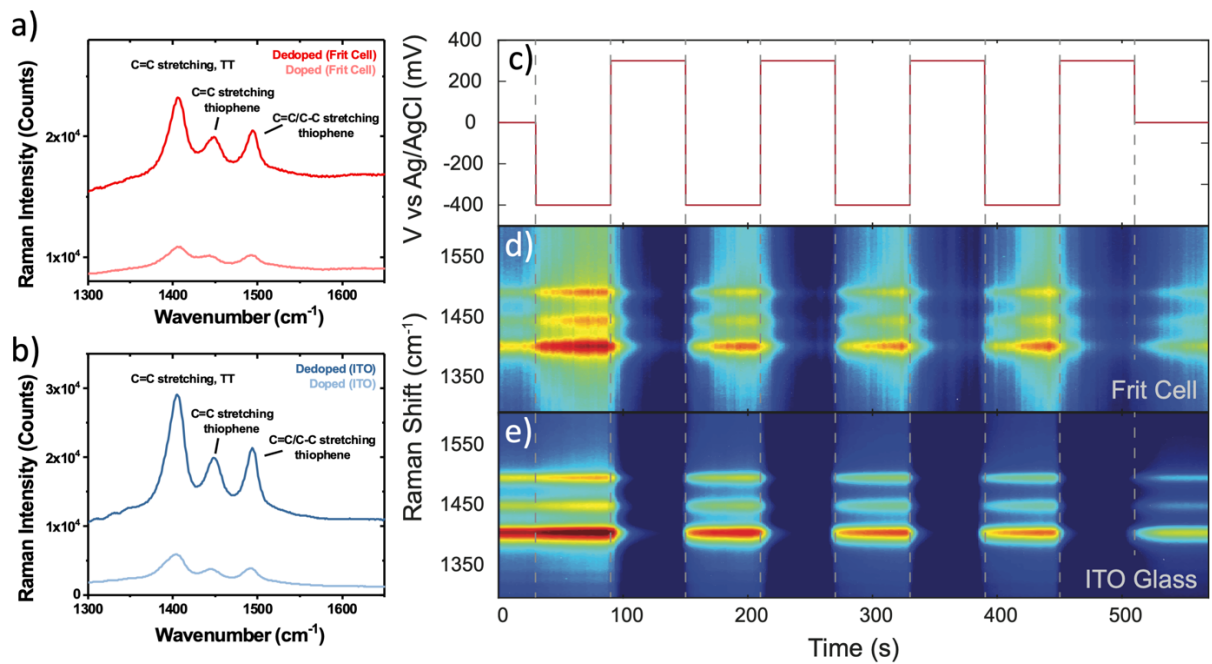

**Figure S4.** Raman Spectroscopy: equilibrated Raman spectra of doped (oxidized) and dedoped (reduced) pg2T-TT on (a) a Au coated frit electrode substrate and (b) an ITO coated glass substrate. (c) potential profile and color plots of Raman spectra of pg2T-TT on (d) a Au coated frit electrode substrate and (e) an ITO coated glass substrate during electrochemical cycling.

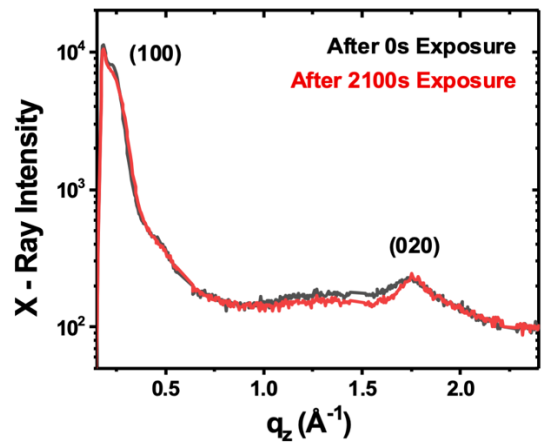

**Figure S5.** Exposure and cycling stability: out-of-plane line cuts from the electrolyte exposed film in the frit based in situ cell when initially exposed and after 2100 s of exposure and electrochemical cycling.